\documentclass[aps,prb,groupedaddress,a4paper,bibnotes,twocolumn,longbibliography]{revtex4-1}
\usepackage{float}
\usepackage{units}
\usepackage{graphicx}
\usepackage{amssymb}
\usepackage{microtype}
\graphicspath{{./figure/}}
\usepackage[charter,greekuppercase=italicized]{mathdesign}

\usepackage{natbib}
\newcommand{\ped}[1]{\ensuremath{_{\rm #1}}}

\usepackage{color}
\usepackage{url}

\usepackage[colorlinks,plainpages=false,linkcolor=blue,urlcolor=blue,citecolor=blue,pdfpagemode=UseNone,pdfstartview=FitBH]{hyperref}

\newcommand{\tcr}[1]{\textcolor{black}{#1}}
\newcommand{\tcb}[1]{\textcolor{black}{#1}}

\begin{document}
\title{Superconductivity of underdoped PrFeAs(O,F) investigated via point-contact spectroscopy and 
nuclear magnetic resonance} 

\author{D.\ Daghero}
\email{dario.daghero@polito.it}
\affiliation{Department of Applied Science and Technology, Politecnico di Torino, 10129
Torino, Italy }

\author{E.\ Piatti}
\affiliation{Department of Applied Science and Technology, Politecnico di Torino, 10129
Torino, Italy }

\author{N.\ D.\ Zhigadlo}
\email{nzhigadlo@gmail.com}
\affiliation{CrystMat Company, CH-8046 Zurich, Switzerland}

\author{G.\ A.\ Ummarino}
\affiliation{Department of Applied Science and Technology, Politecnico di Torino, 10129 Torino, Italy}
\affiliation{National Research Nuclear University MEPhI (Moscow Engineering Physics Institute),
Kashira Hwy 31, Moskva 115409, Russia}

\author{N.\ Barbero, T.\ Shiroka}
\email{tshiroka@phys.ethz.ch}
\affiliation{Laboratory for Muon-Spin Spectroscopy, Paul Scherrer Institut, CH-5232 Villigen PSI, Switzerland}
\affiliation{Laboratorium f\"ur Festk\"orperphysik, ETH Z\"urich, CH-8093 Zurich, Switzerland}

\date{\today}
\begin{abstract}
Underdoped PrFeAs(O,F), one of the less known members of the 1111 family 
of iron-based superconductors, was investigated in detail by means of 
transport, SQUID magnetometry, nuclear magnetic resonance (NMR) 
measurements and point-contact Andreev-reflection spectroscopy (PCARS).
PCARS measurements on single crystals evidence the multigap nature of 
PrFeAs(O,F) superconductivity, shown to host at least two isotropic gaps, 
clearly discernible in the spectra, irrespective of the direction of 
current injection (i.e., along the $ab$ planes or along the $c$ axis). 
Additional features at higher energy can be interpreted as signatures 
of a strong electron-boson coupling, as demonstrated by \tcr{a model which 
combines} Andreev reflection with the Eliashberg theory. Magnetic 
resonance measurements in the normal phase indicate the \tcr{lack of a} 
magnetic order in underdoped PrFeAs(O,F), while $^{75}$As 
NMR spin-lattice relaxation results suggest the presence of significant 
electronic spin fluctuations, peaking above $T_{c}$ and expected 
to mediate the superconducting \tcr{pairing.} 
\end{abstract}


\keywords{Suggested keywords}

\maketitle

\section{Introduction}\label{sec:intro}
The discovery of superconductivity in \textit{Ln}FeAsO oxypnictides 
(\textit{Ln}1111, \textit{Ln}: lanthanide) generated a widespread 
interest among the condensed matter physicists \cite{Kamihara2008}. 
These compounds, which belong to the 1111 family of Fe-based 
superconductors, exhibit a ZrCuSiAs-type structure, composed of 
alternating stacks of \textit{Ln}O and FeAs layers. 
They become superconductors either through chemical substitution 
at the different atomic sites, or through the application 
of external pressure~\cite{Iimura2012,Kuzmicheva2019,Khasanov2011}. 
Consequently, the resulting electronic phase \tcr{diagrams depend}  
sensitively on the particular doping element. In either case, the 
original \tcb{antiferromagnetic} 
state is partially or fully suppressed. 
In particular, it has been shown that the \textit{Ln}FeAsO parent 
compounds can be doped with holes by partially replacing the 
\textit{Ln}$^{3+}$ ions with \tcr{divalent ions,} such as Sr$^{2+}$, 
as e.g.\ in La$_{1-x}$Sr$_{x}$FeAsO~\cite{Wen2008} or 
Pr$_{1-x}$Sr$_{x}$FeAsO \cite{Mu2009}. By converse, 
$n$-type doping can be \tcr{achieved} by substituting \textit{Ln}$^{3+}$ 
with \tcr{tetravalent ions,} such as 
Th$^{4+}$ (Sm$_{1-x}$Th$_{x}$FeAsO)~\cite{Wang2008,Zhigadlo2010}, 
or by partially replacing O$^{2-}$ with F$^{-}$ or H$^{-}$ 
(\textit{Ln}FeAsO$_{1-x}$F$_{x}$, \textit{Ln}FeAsO$_{1-x}$H$_{x}$)~\cite{Iimura2012,Kuzmicheva2019}
In addition, in case of isovalent doping of the \textit{Ln}1111 
parent compound, as e.g., in the As$_{1-x}$P$_{x}$ case, one 
can tune the magnetic interactions without changing the carrier 
concentration~\cite{Wang2009,Zhigadlo2011}. 
Until now, the electron-doped \textit{Ln}1111-type oxypnictides 
(O$_{1-x}$F$_{x}$ and O$_{1-x}$H$_{x}$) \tcr{seem to exhibit} the 
highest $T_{c}$'s. The control of $T_{c}$ through carrier concentration is, 
therefore, a versatile and powerful mean of elucidating the intrinsic 
nature of superconductivity. 

In most \textit{Ln}1111 families, an increase in doping level shifts 
the system from an antiferromagnetically (AF) ordered state towards 
a purely superconducting (SC) state, via a region where the AF and 
SC phases coexist~\cite{Drew2009,Sanna2009,Lamura2015}. 
By contrast, in PrFeAsO$_{1-x}$F$_{x}$, the N\'eel order (and the 
tetragonal-to-orthorhombic structural transition) appear to vanish 
rather rapidly, possibly in a first-order-type transition, as the 
fluorine concentration approaches the critical value 
$x \sim 0.08$~\cite{Rotundu2009}. 
This behavior has been observed to occur also in the 
LaFeAsO$_{1-x}$F$_x$ family~\cite{Luetkens2009}. 
On the other hand, \tcr{it} differs significantly from the structurally 
related families (where the \textit{Ln} ion is, e.g., Sm, Nd, Ce, etc.), 
whose AF-to-SC \tcr{transitions are} much more extended. 
To investigate this in \tcr{further} detail, homogeneously underdoped 
samples, preferentially in a single-crystalline form, are required.

To date, despite extensive evidence that superconductivity in 
Fe-based materials is mediated by spin fluctuations, a conclusive 
experimental confirmation is still missing. In particular, 
the interplay between the AF fluctuations and superconductivity in 
the underdoped regime remains unclear, mainly reflecting the 
difficulties associated with the preparation of high-quality 
underdoped \textit{Ln}1111 samples. 
The first step toward the elucidation of the nature of superconductivity 
is the growth of high-quality crystals.
 
Here, we report on advanced point-contact Andreev reflection 
spectroscopy (PCARS) and nuclear magnetic resonance (NMR) studies of 
superconductivity in underdoped PrFeAsO$_{1-x}$F$_{x}$ crystals with 
\tcr{a} $T_{c}$ of \tcr{$\sim 24$\,K.} 
Through an exhaustive set of measurements, we directly \tcr{assess} the 
multigap nature of superconductivity in this compound and \tcr{determine} 
the amplitudes of the gaps, that appear to be isotropic in-plane and 
out-of plane, with no evidence of extended node lines. We \tcr{bring} 
evidence of \tcr{a} strong coupling between electrons and a bosonic mode, 
whose characteristic energy agrees well with that of spin fluctuations. 
Finally, we \tcr{show} that the magnetic order, typical of the parent 
compound\tcr{s,} is completely suppressed in these underdoped crystals, while 
sizable spin fluctuations persist, as indicated by NMR. Altogether, 
these results strongly point towards a spin-fluctuation mediated 
multiband superconductivity in PrFeAsO$_{1-x}$F$_{x}$.

\section{Crystal growth and experimental details}\label{sec:growth}
The PrFeAs(O,F) crystals were grown by using a cubic-anvil high-pressure 
high-temperature technique. The details of the setup can be found 
in Ref.~\onlinecite{Zhigadlo2012,Zhigadlo2013}. Starting powders of 
PrAs, FeF$_{2}$, Fe$_{2}$O$_{3}$, and Fe of high purity 
($\geq 99.95$\%) were weighed according to the stoichiometric ratio, 
thoroughly grounded in a mortar and then mixed with NaAs flux.
For one growth batch we used 0.45\,g of PrFeAsO$_{0.60}$F$_{0.35}$ and 
0.2\,g of NaAs. 
%
%
The crystal growth process was performed by heating the mixture
up to $\sim 1500^{\circ}$C in 2\,h. The mixture was kept there 
for 5\,h, cooled to 1250$^{\circ}$C in 60\,h, held at this temperature 
for 3\,h, and finally cooled down to room temperature. The crystalline 
products were separated by dissolving the flux in distilled water. 
Further details on the crystal growth of PrFeAs(O,F) can be found in 
Ref.~\onlinecite{Zhigadlo2013}.

The x-ray analysis confirmed that the obtained crystals belong to 
the 1111-type structure, with the refined \tcr{model} being 
consistent with that from our previous x-ray diffraction 
studies (see Table~1 in Ref.~\cite{Karpinski2009}). 
\tcr{Compositional analysis via} energy-dispersive x-ray (EDX) measurements 
confirmed that the ratio of praseodymium, iron, and arsenic is close to 1:1:1. 
Light elements such as oxygen and fluorine cannot be measured 
accurately via EDX. Therefore, we could not determine the exact doping 
level of the PrFeAs(O,F) crystals. Nevertheless, by a comparison of our 
transition temperatures with those of polycrystalline 
samples~\cite{Rotundu2009} (see below), we estimate an F doping of 
$\sim 0.1$ in our case.

The details of the PCARS technique are given in App.~\ref{app:PCAR_basics}. 
As for the NMR study, this consisted in $^{75}$As lineshape- and 
spin-lattice relaxation measurements, performed at 7.057\,T over a 
temperature range from 4 to 295\,K.
The NMR signals were detected by means of standard spin-echo sequences, 
consisting in $\pi/2$ and $\pi$ pulses of 3 and 6\,$\mu$s, with recycling 
delays ranging from 0.01 to 1\,s. The lineshapes were obtained via 
fast Fourier transform (FFT) of the echo signal. Spin-lattice relaxation 
times $T_1$ were measured via inversion recovery, by using a 
$\pi$--$\pi/2$--$\pi$ pulse sequence.

\section{Superconducting properties}
\subsection{Preliminary characterization of superconductivity}
\begin{figure}
\includegraphics[width=0.8\columnwidth]{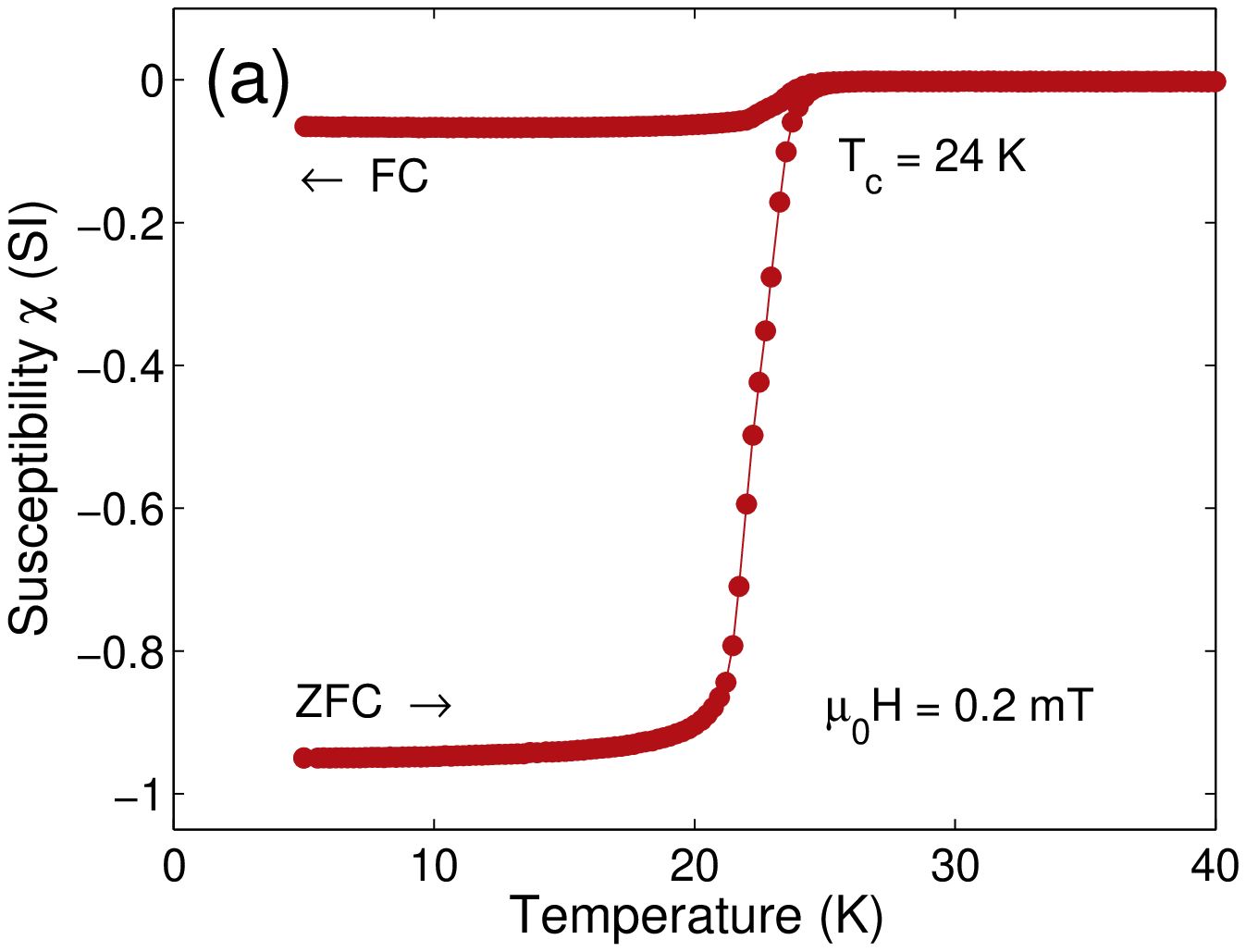}
\includegraphics[width=0.8\columnwidth]{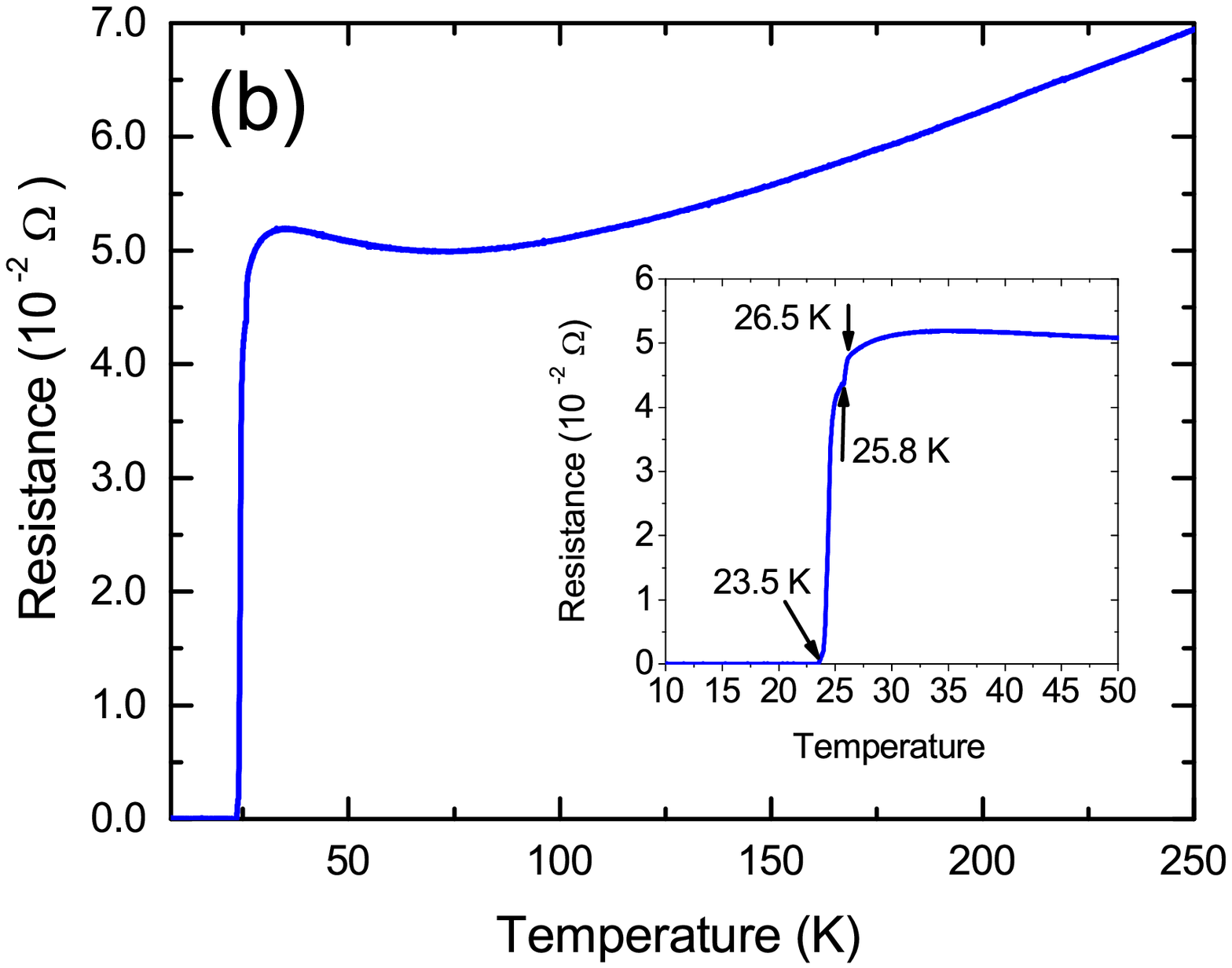}
\caption{(a) Temperature dependence of magnetic susceptibility of 
a \tcr{single} PrFeAs(O,F) crystal. The zero-field-cooled (ZFC) curve was 
obtained on heating in a magnetic field of 0.2\,mT applied along the 
$c$ direction. (b) Resistance 
as a function of temperature. The inset shows a closeup of the 
superconducting transition.} 
\label{fig:magn_res}
\end{figure}

The dependence of magnetic susceptibility vs.\ temperature in 
a \tcr{single} PrFeAsO$_{1-x}$F$_{x}$ crystal, measured in a magnetic field of 
0.2\,mT parallel to the $c$ axis, is shown in Fig.~\ref{fig:magn_res}(a). 
Here, the effective superconducting transition temperature $T_{c,\mathrm{eff}}$ 
is defined as the crossing of the linear extrapolations from the two regions 
of the high-temperature normal state and low-temperature superconducting state. 
In the underdoped case the transition is relatively sharp, indicative of 
a good sample quality.

The resistance was measured by using a standard four-probe technique, 
with the current flowing in the $ab$ plane.
Upon lowering the temperature, as shown in Fig.~\ref{fig:magn_res}(b), 
the resistance first decreases linearly, to reach a broad minimum around 
70\,K, and then it increases again. Similar features were also observed 
in the other investigated crystals. 
A closeup of the superconducting transition region is shown in the 
inset. After a saturation around 30\,K the resistance starts \tcr{dropping} 
and reaches \tcr{its} zero value at 23.5\,K, fully consistent with the onset of 
the magnetic transition as measured via SQUID magnetometry.
This behavior (and the relevant values) are similar to those observed 
in polycrystals with $x=0.11$ by Rotundu et al.~\cite{Rotundu2009}. 
A very tiny kink at 25.8\,K might indicate the presence of another 
superconducting phase with a close lying $T_c$ value and, hence, 
with practically the same F doping. 
It is interesting to note that there seems to be a connection 
between the position of the minimum in the normal-state resistance 
and the critical temperature $T_c$ (i.e., the two differ by a factor 
of about 3 in all the measured crystals).


\subsection{Superconducting energy gaps}
Further insight into the superconducting properties of PrFeAs(O,F) 
is obtained from the point-contact Andreev reflection spectroscopy, 
which allows us to directly investigate the superconducting gap structure. 
The technique is quite simple and consists in measuring the differential 
conductance, $dI/dV$, of a pointlike contact between a normal metal and 
the superconductor under study, as a function of the bias voltage $V$ 
across the contact (see App.~\ref{app:PCAR_basics} for details). 

\begin{figure}
\includegraphics[width=0.9\columnwidth]{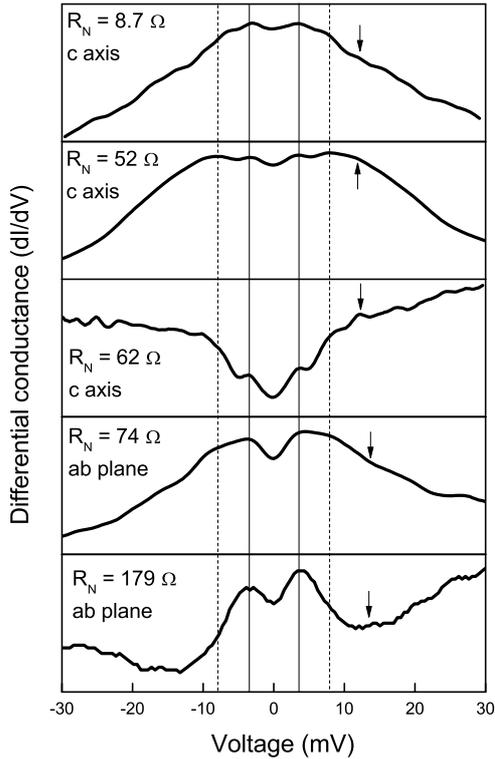}
\caption{Low-temperature (2.7\,K) conductance curves for various contacts 
made on the same single crystal. 
Despite the different shapes, all the curves display structures (maxima, shoulders, slope changes) at approximately the same positions, as indicated by the 
vertical lines at $\pm 3.5$\,mV and $\pm 7.5$\,mV.  
Arrows indicate structures related to the 
strong electron-boson coupling.} 
\label{fig:lowTconductance}
\end{figure}

Figure~\ref{fig:lowTconductance} shows typical conductance curves, 
measured at 2.7 K, for contacts made 
either on the top surface ($c$-axis contacts) or on the side ($ab$ plane contacts). The labels refer to the direction of current injection, i.e., perpendicular or parallel to the FeAs planes, respectively. Despite the different shapes 
of the curves, the different directions of current injection, and the 
different resistance of the contacts, it is clear that they all show 
structures at approximately the same energies. In particular, the position of the low-energy maxima ($\pm 3.5$ meV, solid vertical lines) is very robust. Additional features, that can take the form of maxima, shoulders, or slope changes are present at about $\pm 7.5$ mV (dashed vertical lines). These two values are particularly interesting because a full SC gap with an amplitude $1.6 k_\mathrm{B}T_c = 3.5$\,meV was observed in underdoped PrFeAs(O,F) single crystals by microwave penetration depth and by quasiparticle conductivity measurements \cite{hashimoto2009}, 
while a gap of $3.5 k_\mathrm{B} T_c \simeq 7.5$\,meV was detected by optical conductivity measurements \cite{charnukha2018}. Other structures whose occurrence is apparently less systematic can be observed at higher energies (arrows). From the spectroscopic point of view, the fact that there are structures whose position does not depend on the resistance of the contacts means that: i) all the contacts are spectroscopic, at least at low temperature; ii) these structures are intrinsic, i.e., unrelated to the contact, but instead directly connected to the properties of the material. In particular, they are suggestive of the presence of multiple (at least two) superconducting energy gaps. This is a rather common feature of Fe-based systems (including here materials of the same family, such as La-1111 \cite{gonnelli2009} and Sm-1111 \cite{daghero2009}). 
It is worth noting also that the position of the spectral features 
does not depend on the direction of current injection, which suggests that the system does not show a clear in-plane/out-of-plane anisotropy, at least for the gap amplitudes. Clearly, one cannot exclude small anisotropies (i.e., $k$ 
dependence of the SC gaps), undetectable by our technique.

\begin{figure}
\includegraphics[width=0.9\columnwidth]{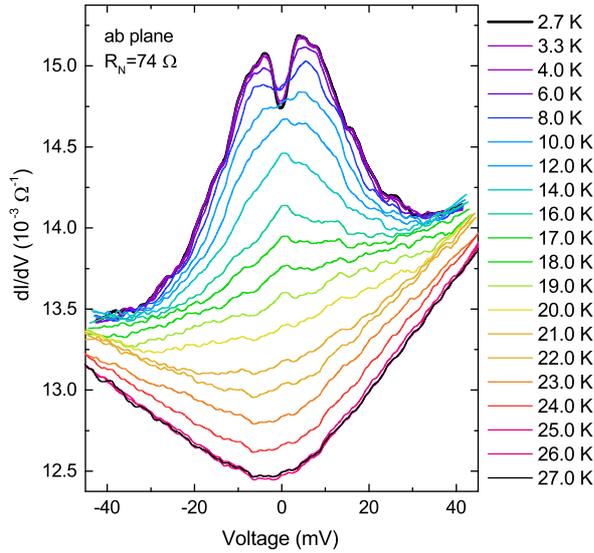}
\caption{Temperature dependence of the conductance curves of a 
74-$\mathrm{\Omega}$, $ab$-plane contact on a PrFeAs(O,F) single crystal. 
Upon increasing $T$, the Andreev-reflection features decrease in amplitude  and disappear between 24 and 25\,K.}\label{fig:Tdep}
\end{figure}

Finally, the shape of the curves and, in particular, the absence of zero-bias maxima indicates the absence of significant contributions of low-energy quasiparticles to the conductance, thus suggesting a fully-gapped SC and the absence of node lines (as also demonstrated by microwave penetration depth measurements in underdoped single crystals \cite{hashimoto2009}). A similar situation was observed in other 1111 compounds like F-doped Sm-1111 and La-1111 \cite{daghero2009,gonnelli2009}. By contrast, the PCARS spectra of some 122 systems, featuring accidental node lines, show zero-bias maxima at least in one of the directions of current injection; a typical example being the Ca-122 system \cite{gonnelliSciRep2016,gonnelli2012}. Based on the spectra shown in 
Fig.~\ref{fig:lowTconductance}, and consistently with the results of penetration depth \cite{hashimoto2009}, critical field \cite{hashimoto2009}, and infrared spectroscopy \cite{charnukha2018} measurements, as well as with electronic 
band-structure calculations for 1111 compounds, from now on we will 
assume PrFeAs(O,F) to be a multiple-gap, $s\pm$-wave superconductor.

Figure \ref{fig:Tdep} shows the conductance curves of a 74-Ohm, $ab$-plane 
contact on a PrFeAs(O,F) single crystal as a function of temperature. The lowest-temperature spectrum was already depicted in the second-last panel of 
Fig.~\ref{fig:lowTconductance}. The temperature dependence is crucial 
in identifying the critical temperature $T_c^A$ where the Andreev signal 
disappears, thus allowing us to determine the normal-state conductance of the contact. As a matter of fact, the progressive decrease in amplitude of the spectra on increasing temperature is due to the decrease in amplitude of the superconducting gap(s). Hence, $T_c^A$ can be identified with the 
temperature at which the curves recorded at increasing temperatures 
start to overlap 
and the Andreev-reflection features disappear. In our case, the gap vanishes between 24\,K and 25\,K, since the 25-K conductance curve is 
superimposed to those recorded at 26\,K and 27\,K. 
Thus, the curve recorded at 25\,K represents the normal-state spectrum of the contact and we will assume $T_c^A = 24.5 \pm 0.5$\,K. Note that this value is consistent with the $T_c$ obtained from magnetometry and transport data  
(see Fig.~\ref{fig:magn_res}). This, again, is an indirect, yet quite 
convincing proof of the spectroscopic nature of the contact. Indeed, if the conduction through the contact were 
diffusive, Joule heating would occur \emph{within the contact} \cite{daghero2010} and the Andreev signal would disappear at a lower bath temperature.


\begin{figure}
	\includegraphics[width=0.9\columnwidth]{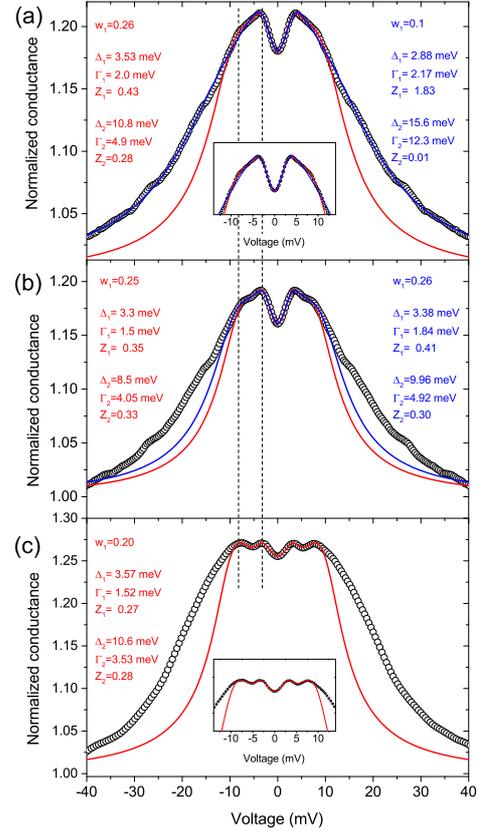}
	\caption{(a) Open circles: low-temperature (2.7\,K) conductance data  
	of a 74-$\mathrm{\Omega}$, $ab$-plane contact on a PrFeAs(O,F) single 
	crystal,  
	 after normalization (i.e., division by the normal-state conductance curve without any shift). The blue line is a two-gap fit of the whole curve. The red curve is a fit of the central part, excluding the wide shoulders associated to the electron-boson coupling structures. (b) Same curve as in (a), but with a different normalization (division by the normal-state conductance curve shifted upwards). The two fitting curves were obtained with different values of $V_\mathrm{max}$. (c) Normalized low-temperature conductance data of a $c$-axis, 52-$\mathrm{\Omega}$ contact (circles) with the relevant best fit. Vertical dashed lines highlight the correspondence of the position of the gap structures.}\label{fig:lowTfit}
\end{figure}

To extract the gap amplitudes more accurately, the conductance curves 
must first be normalized and then fitted to a suitable model.
The normalization is obtained by dividing the differential conductance recorded at a given $T<T_c$ by the normal-state conductance of the same contact. The curve measured just above $T_c$ can be used, under the reasonable assumption that the normal-state properties do not change much between $T$ and $T_c$. The lowest-temperature curve in Fig.~\ref{fig:Tdep}, once divided by the normal-state curve recorded at 27\,K (and symmetrized, to better highlight the intrinsic structures and to suppress noise fluctuations) is shown in Fig.~\ref{fig:lowTfit}(a) (circles). On top of it, we plot some theoretical curves, obtained through an automatic fitting procedure based on the minimization of the sum of squared residuals. The model used to fit the experimental data 
is a two-band, 2D version of the Blonder-Tinkham-Klapwijk \tcb{(BTK)} model \cite{BTK,kashiwaya,daghero2010}. This model contains as free parameters the amplitude of the gaps $\Delta_1$ and $\Delta_2$, the broadening parameters $\Gamma_1$ and $\Gamma_2$, the barrier parameters $Z_1$ and $Z_2$ and the relative weight of band 1 in the conductance, $w_1$. These parameters are not completely free, 
because the values of the gaps reflect 
the position of the maxima and shoulders, and the barrier parameters 
determine the percentage of tunnel vs.\ Andreev-reflection conduction 
through the junction and, in practice, are related to the depth of the zero-bias minimum and to the shape of the curve between $\Delta_1$ and $\Delta_2$. Details of the fitting procedure can be found elsewhere \cite{daghero2010}.
The blue curve in Fig.~\ref{fig:lowTfit}(a) is an attempt to fit the conductance 
data across the whole voltage range. The overall fit seems fairly good, yet the fitting function completely fails to reproduce the features at $\pm 7.5$\,mV, 
most likely reflecting a superconducting gap (see inset). Moreover, the amplitude of the large gap $\Delta_2= 15.6$\,meV is far too big for 
a system with $T_c = 25$\,K, since the gap ratio $2\Delta_2/k_\mathrm{B} T_c$ would be 14.5. This value is completely unreasonable even though, in other compounds of the 1111 family, the (larger) gap ratio can be as high as 8 \cite{daghero2009,gonnelli2009}. Finally, the values of the $\Gamma$ parameters are too high, and comparable to the gap values themselves, which should not happen in a  spectroscopic contact.
The fit is thus unsatisfactory and meaningless. The reason is that, as already demonstrated in the case of Ba(Fe,Co)$_2$As$_2$ \cite{tortello2010}, SmFeAs(O,F) \cite{dagheroRopp2011}, and Fe(Te,Se) \cite{daghero2014}, in Fe-based superconductors the relatively strong coupling between the electrons and 
bosons that mediates the Cooper pairing gives rise to additional structures (shoulders) in the tunnel- and PCARS spectra, better seen as peaks in the second derivative $-d^2I/dV^2$, that do \emph{not} occur at the gap edge, but at a higher energy $E_p$. \tcb{As discussed in Ref.~\cite{dagheroRopp2011}, in case of multiple gaps this energy is $E_p\simeq \Delta\ped{max}+\Omega_0$, where $\Omega_0$ is the characteristic boson energy and $\Delta\ped{max}$ is the largest gap. The electron-boson interaction does not affect the spectra in the energy region where the gap features are observed, but it gives rise to shoulders that can extend to rather high energies and can enormously enhance the apparent width of the conductance curve. These structures} \emph{cannot} be fitted by the Blonder-Tinkham-Klapwijk model, even in its various extended versions, if energy-independent gaps are used as in the BCS, weak-coupling theory. To include the effects of strong coupling in the theory, a much more complicated procedure has to be used, which involves the solution of the Eliashberg equations (see Appendix \ref{app:PCAR_theory}). \tcb{As demonstrated in Appendix \ref{app:PCAR_theory}, since the BCS theory represents the low-energy limit of the Eliashberg theory, the low-bias region of the spectrum is completely and uniquely determined by the Andreev reflection. Hence, in this region, the BTK model with constant gaps can be safely used to extract the gap values \cite{Parks}.} In Fe-based compounds, the superconductivity is thought to be mediated by spin fluctuations. Indeed, the position of the electron-boson structures we observe in the aforementioned materials agrees well with a characteristic boson energy $\Omega_0$ that obeys the empirical law $\Omega_0 \simeq 2T_c/5$, where $T_c$ is in kelvin and $\Omega_0$ in meV \cite{Inosov}. In our samples, $\Omega_0 \simeq 10$ meV and the structures are expected to fall at energies larger than the maximum gap amplitude, that we will call $\Delta_2$.

We have thus to abandon the idea of fitting the \emph{whole} curve, and 
focus instead on the low-energy region that hosts the structures related to the gaps, i.e., on the region $|V| \leq V_\mathrm{max}$. The choice of $V_\mathrm{max}$ is somewhat arbitrary and can (slightly) affect the values of the energy gaps. The red curve in Fig.~\ref{fig:lowTfit}(a) was obtained by setting $V_\mathrm{max}=10$\,mV, which implies much more reasonable values for the parameters (reported in the labels). In particular, the value of the small gap $\Delta_1=3.53$ meV is perfectly compatible with the results of penetration depth and theryarticle conductivity measurements \cite{hashimoto2009}.

The fact that the high-energy tails of the unnormalized curves (Fig.~\ref{fig:Tdep}) are affected by the electron-boson structures, and the fact that these structures \emph{depend} on the energy gap and disappear only at $T_c$ \cite{tortello2010,dagheroRopp2011}, means that also the normalization is somewhat arbitrary. The usual criterion, i.e., that the high-voltage tails ($V>3\Delta_2$) of the conductance curves must fall on top of the normal-state conductance, is no longer true. To be conservative, we thus tried different normalizations, obtained by vertically shifting the normal-state conductance by different amounts. We found that the amplitude of the small gap is very robust, being determined by the energy position of the maxima, while the value of the large gap depends somewhat on the height of the normalized curve, that in turn depends on the normalization. For example, Fig.~\ref{fig:lowTfit}(b) reports the same curve, with a different normalization (i.e., divided by the normal-state conductance shifted slightly  upwards) with two fits, obtained by using different values of $V_\mathrm{max}$. We will keep trace of this variation by using proper error bars on the gap values.

\begin{figure}
	\includegraphics[width=0.88\columnwidth]{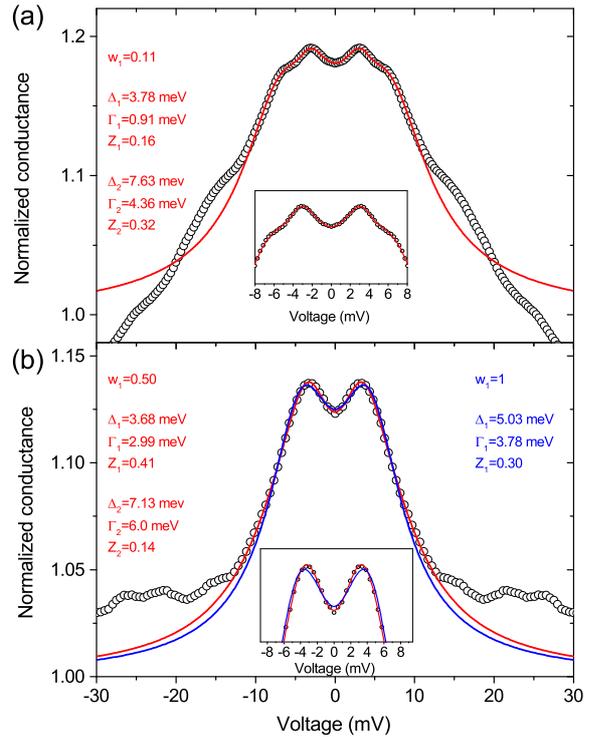}
	\caption{(a) Normalized differential conductance at 2.7\,K (circles) and relevant fit (red line) for a $c$-axis contact with a resistance of 8.7\,$\mathrm{\Omega}$.  (b) Open circles: low-temperature (2.7\,K) conductance spectrum of a 179-$\mathrm{\Omega}$, $ab$-plane contact, 
		divided by the normal-state conductance curve without any shift. The blue line is a single-gap fit, while the red curve 
		is a two-gap fit, both made with $|V_\mathrm{max}|=10$\,mV. The inset shows a closeup of the low-bias region.}\label{fig:lowTfit2}
\end{figure}

Figure~\ref{fig:lowTfit}(c) shows the normalized low-temperature conductance curve of a 52-$\mathrm{\Omega}$, $c$-axis contact. As in the previous case, the curve presents maxima around 3\,mV and structures at $\sim 7.5$\,meV, that here appear as clear maxima. The fit, which disregards the shoulders clearly associated to the electron-boson structures, gives a small gap $\Delta_1=3.57$\,meV (in perfect agreement with what found in the previous case) and a large gap $\Delta_2 = 10.6$\,meV, whose value exhibits a certain variability, depending on the normalization. In the spectrum shown in \tcb{Fig.~\ref{fig:lowTfit2}(a)}, recorded at 2.7\,K in a $c$-axis contact with resistance 8.7\,$\mathrm{\Omega}$, the features associated to the large gap are rather clear and well separated from the electron-boson structures. Here, the best fit gives $\Delta_1=3.78$\,meV and $\Delta_2 = 7.63$\,meV.

\tcb{Figure~\ref{fig:lowTfit2}(b) shows the low-temperature conductance curve of an $ab$-plane point contact on the same crystal. Here, the features related to the small gap are dominant and no clear structures associated to the large gap can be detected by eye. With the normalization shown in the figure (obtained without any shift of the normal-state conductance), the two-gap fit (red line) is superior to the single-gap one (blue line), because it can reproduce both the position of the maxima (see inset) and the width of the curve. This fit gives $\Delta_1 = 3.68$\,meV and $\Delta_2=7.13$\,meV. The gap value obtained by the single-band fit is $\Delta=5.03$\,meV and can be seen as a sort of an average of $\Delta_1$ and $\Delta_2$, as usually happens when multiple gaps are insufficiently resolved in the spectra. Actually, by choosing different normalizations, the single-band fit can become almost indistinguishable from the two-band one (in particular, if the amplitude of the normalized curve is lowered) and it always provides a gap $\Delta$, intermediate between $\Delta_1$ and $\Delta_2$. Therefore, on the basis of this spectrum alone, one cannot decide between the single-band and the multi-band picture. However, the two-gap picture is compatible with all the other spectra, and there is no reason to expect a different behaviour in this particular case.}


\begin{figure}[ht]
	\includegraphics[width=\columnwidth]{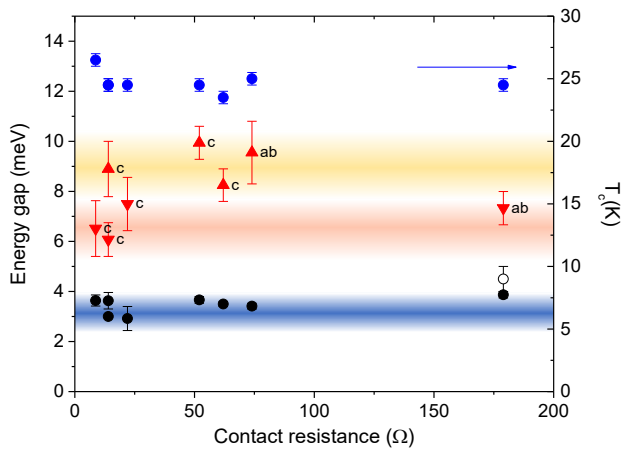}
	\caption{Energy gaps (red and black solid symbols) and Andreev critical temperature (blue symbols) as a function of the contact resistance. \tcb{The open symbol represents the result of a single-gap fit of the curve shown in Fig.~\ref{fig:lowTfit2}(b)}. The labels indicate the type of contact (i.e., with current parallel to the $c$ axis or to the $ab$ planes). }
	\label{fig:gaps}
\end{figure}

Figures~\ref{fig:lowTfit} and \ref{fig:lowTfit2} are representative of 
all the datasets we collected: while the value of the small gap is 
always the same (within experimental uncertainties), the values of the large gap seem to fall into two energy intervals: one centered around 7\,meV 
and another around 10\,meV. \tcb{In principle, this clustering could indicate that there are actually \emph{two} different large gaps. The fact that we detect either one or the other may be due to the use of a \emph{two}-gap fit function (a model with three gaps would have 12 free parameters, too many for the fit procedure to be meaningful). The presence of more than two gaps is not unusual for Fe-based superconductors, considering that in these compounds three or more bands cross the Fermi level.} 


Figure~\ref{fig:gaps} summarizes the values of the gaps extracted from the fit of all the conductance datasets, as a function of the contact resistance. The value of the small gap is very well defined and completely independent on the resistance, which is the last and definitive proof that the contacts are spectroscopic. The values of the large gap(s) are more scattered and affected by the uncertainty arising from the aforementioned degree of freedom introduced by the electron-boson structures. These, although excluded from the fits, can still affect the normalization.
The figure highlights also the energy range where the SC gaps occur. As for the large gap, we indicate here the possible ranges in the hypothesis that \emph{two} large gaps are present, although difficult to discern. Only detailed ARPES measurements can possibly disentangle the 
energy gaps residing on each sheet of the Fermi surface.

\section{$^{75}$A\lowercase{s} NMR results: Role of spin fluctuations}
While PCARS provides detailed information on the nature and value of 
the superconducting gaps, NMR can be used to investigate also the 
normal-state properties of PrFeAs(O,F). 
Here, we employ mostly ${}^{75}$As-NMR measurements at 7\,T 
to determine the static (line widths and -shifts), as well as 
the dynamic (spin-lattice relaxation) electronic properties 
of PrFeAs(O,F). The very small size of the single crystals implied 
a rather poor S/N ratio. Therefore, the NMR measurements had to be performed 
on powder samples (obtained by crushing the available single crystals).

\begin{figure}
\includegraphics[width=0.9\columnwidth]{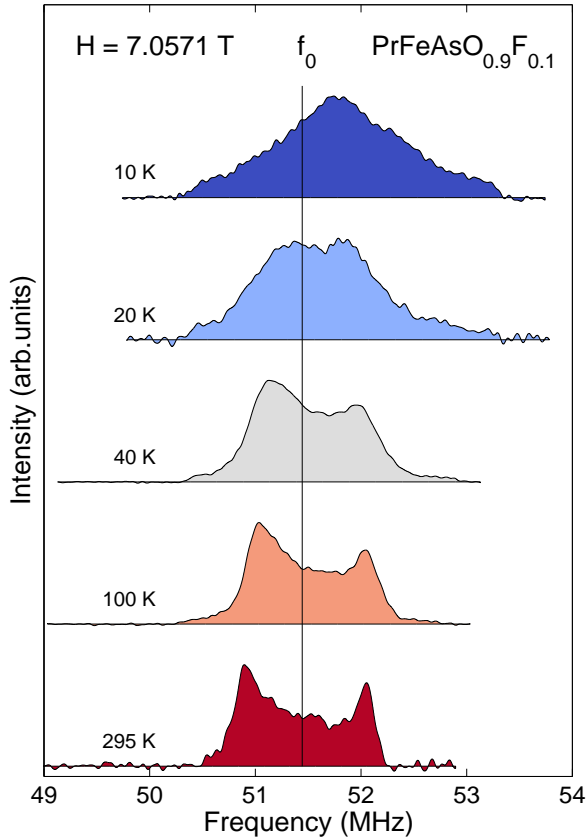}
\caption{Frequency-swept ${}^{75}$As NMR spectra in PrFeAs(O,F), 
measured at 7\,T. The lines refer to the ${}^{75}$As central 
transition, whose features reflect a second-order quadrupole broadening. 
While \tcr{down to $T_c$} the line width is relatively insensitive to temperature, 
its position shifts towards higher frequencies (especially below 
$T_{c}$). The vertical line indicates the reference ${}^{75}$As-NMR 
frequency.}
\label{fig:powder_lines}
\end{figure}

Since ${}^{75}$As has a nuclear spin $I = 3/2$ with a moderately large 
quadrupole moment ($Q = 31.4$\,fm$^{2}$), the observed NMR line consists 
of the central Zeeman  $+$\nicefrac{1}{2}\ to $-$\nicefrac{1}{2}\ 
transition broadened by a second-order quadrupole perturbation, while 
the two satellites are much too weak and far apart. 
Both the two-peak lineshape (see Fig.~\ref{fig:powder_lines}) and its 
variation with temperature \tcb{(at least down to 40\,K) are similar} 
to those of the ${}^{75}$As NMR lines observed in lightly F-doped LaFeAsO 
\cite{Grafe2008,Nakai2009} or in pure ThFeAsO \cite{Shiroka2017}. 

\subsection{NMR lineshapes and lack of magnetic order}
As shown in Fig.~\ref{fig:powder_lines}, down to $T_c$ \tcb{($\sim 15$\,K at 7\,T),} 
the position of 
the NMR lines does not change significantly with temperature, hence suggesting 
the absence of a magnetic order. This is in agreement with the magnetometry 
results which, above $T_c$, also show a weak and almost flat response.
Below $T_c$, instead, the NMR lines do not exhibit the expected drop in 
frequency in the superconducting phase. Indeed, we find that the lines 
maintain their position, or even show a slight increase in frequency 
(see, e.g., the dataset at 10\,K). 
This puzzling behavior has been observed also in other iron-based 
superconductors \cite{Mukuda2014,Shiroka2018}, with a possible 
explanation invoking the multiband structure of such materials. 
In fact, the Knight-shift part due to the electronic spins can be 
decomposed into two components: $K_\mathrm{spin} = A_{s} \chi_{s} 
+ A_{\mathrm{cp}} \chi_{\mathrm{non-s}}$, 
with $A_{s}$ the direct Fermi-contact hyperfine coupling to the 
$s$-electrons and $A_{\mathrm{cp}}$ arising from the core polarization 
of the inner $s$-shells due to non-$s$ (e.g., $p$ or $d$) electrons 
\cite{Abragam1978}. 
If $\chi_\mathrm{non-s}$ is strongly temperature dependent, 
$K_\mathrm{spin}$ \tcr{changes} accordingly and, possibly, even 
reverses its sign.
The multiband structure of 1111 compounds makes things even more 
complicated since, in this case, the spin susceptibility from each 
band may exhibit a different temperature response, depending on the 
overlap \tcr{of the core-} with $p$-orbitals of the As ion. 
\tcb{In addition, in our case, the increasing influence of the magnetic 
Pr$^{3+}$ ions at low temperature might also explain the observed 
increase in frequency below ca.\ 20\,K. Indeed, significant $^{75}$As 
shifts of ca.\ +1.5\% have been observed also in single-crystal NMR 
studies of CeFeAsO$_{0.8}$F$_{0.2}$, in particular for $H \parallel c$ 
\cite{Rybicki2013}.}

As for the NMR linewidths, these too are similar to those of analogous 
compounds (see references above), \tcb{mostly} in terms of the FWHM 
value (here typically around 1.30\,MHz), \tcb{but less in terms of its}  
temperature dependence. 
In the PrFeAs(O,F) case, too, we observe a moderate increase of 
FWHM with decreasing temperature (indicative of enhanced magnetic spin 
fluctuations). \tcb{However, unlike the generic case, here the two-peaked 
$^{75}$As central transition shows a progressive broadening of the peaks, 
which at the same time become closer, until they merge below ca.\ 15\,K. 
While these contrasting trends leave the global FWHM practically unchanged, 
the clear change in line shape indicates a progressive enhancement of the 
Pr$^{3+}$ magnetic effects at low temperature, reflected also in the $1/T_{1}$ 
relaxation rates (see below).}
Since the observed increase of FWHM is smooth, this is in sharp contrast with 
the abrupt changes expected in case of a magnetic phase transition 
\cite{Borsa2007}. 
The lack of appreciable variations of FWHM vs.\ $T$ strongly 
suggests that PrFeAs(O,F) does not exhibit any AF order but, at the 
same time, it may sustain AF fluctuations, as we show in detail below. 
Finally, note that, in our case, the linewidth broadening may also 
arise from quadrupole effects, mostly reflecting disorder or defects 
intrinsic to doped samples. However, comparisons of pure-NQR with NMR spectra 
have shown that the former is of 
secondary importance and does not lead to the observed temperature 
dependence \cite{Baek2012}. In particular, the quadrupole broadening 
is quantitatively less pronounced for the central $+$\nicefrac{1}{2}\ 
to $-$\nicefrac{1}{2}\ transition, affected only to second order by the 
quadrupole effects.

\subsection{NMR relaxation rates and spin fluctuations}
$^{57}$Fe M\"{o}ssbauer spectroscopy studies on the PrFeAsO parent 
compound found an itinerant $3d$ magnetic order of Fe$^{2+}$ ions, with 
an onset at about 165\,K, accompanied by an orthorhombic distortion of 
the unit cell \cite{Komedera2017}. 
Upon lowering the temperature, this evolves into a complete longitudinal 
incommensurate spin-density-wave (SDW) order below 139\,K. At much 
lower temperatures (12.8\,K) also the localized Pr$^{3+}$ magnetic ions 
order. Although the critical temperature of this second magnetic system 
depends on the type of rare-earth \cite{Sanna2009}, such phenomenology 
is common to many 
1111 parent compounds.
Upon F doping, the SDW magnetism is expected to vanish, yet the spin 
fluctuations to survive. Indeed, as we show below, in the lightly doped 
PrFeAs(O,F) case, the $^{75}$As NMR spin-lattice relaxation 
shows clear signatures of strong magnetic fluctuations \cite{Ishida2011}.

\begin{figure}
\centering
\includegraphics[width=0.9\columnwidth]{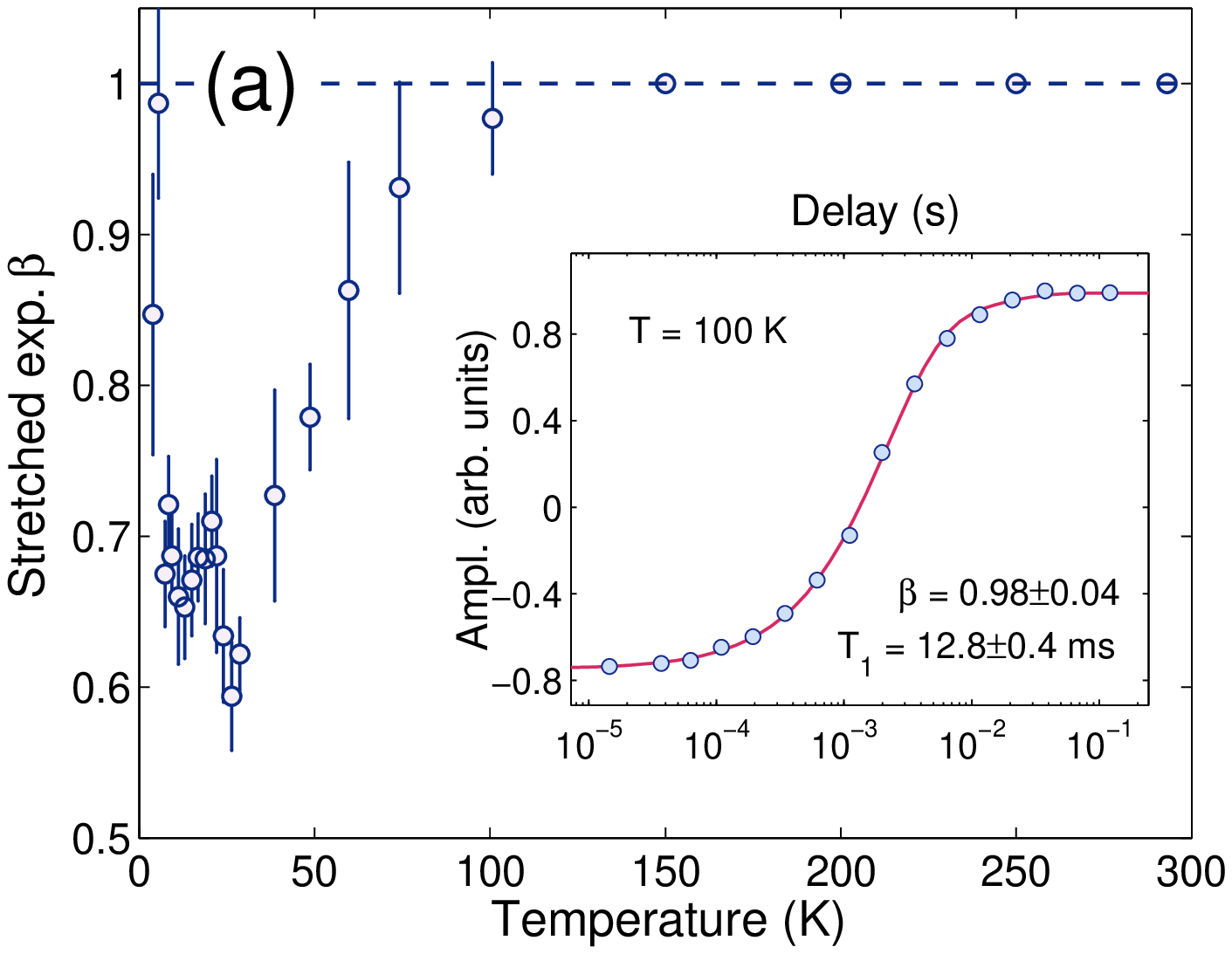}
\hspace*{-2mm}\includegraphics[width=0.93\columnwidth]{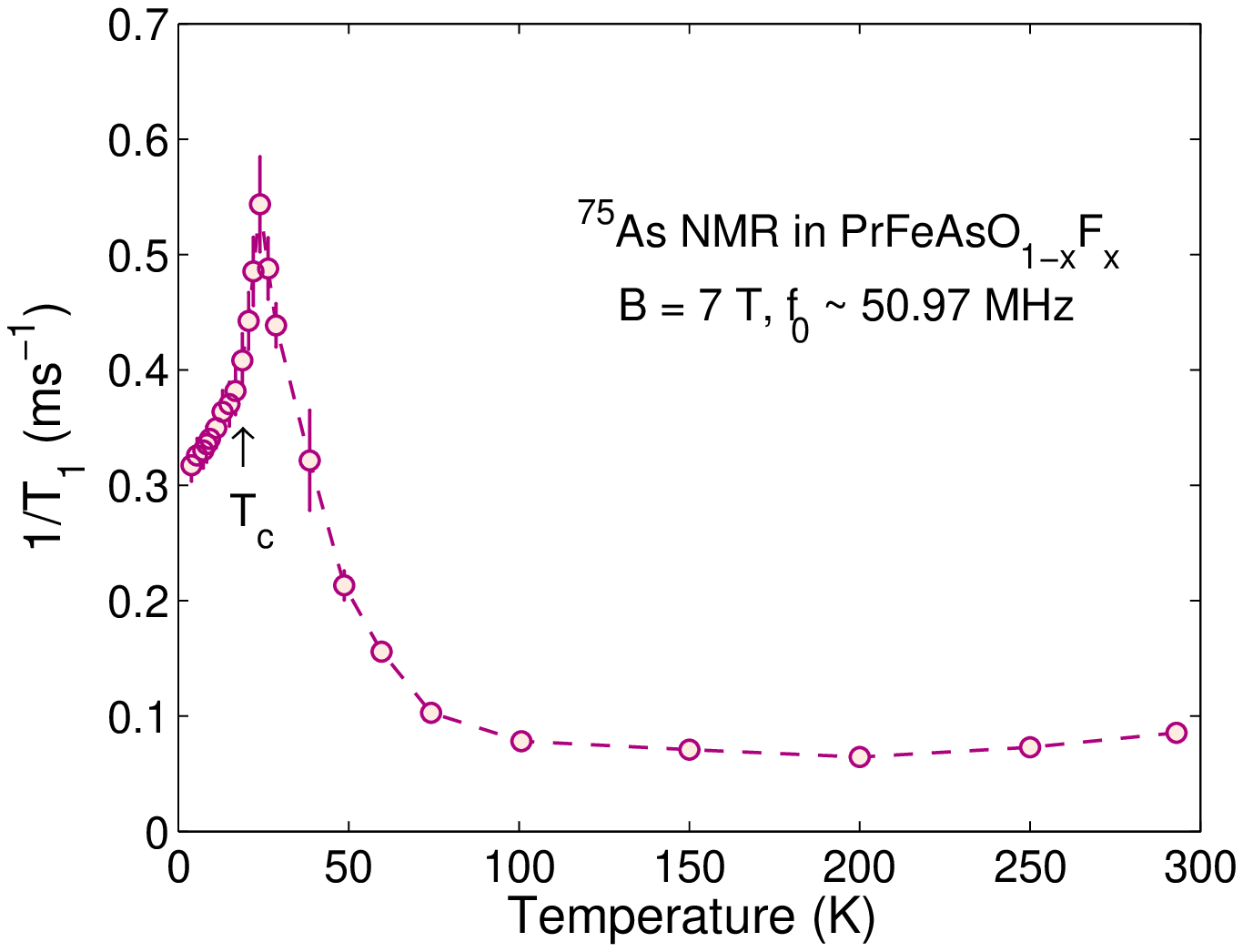}
\caption{(a) Stretched-exponential coefficient $\beta$ vs.\ temperature. 
Below 100\,K, $\beta$ deviates increasingly from 1 (dashed line). 
Inset: recovery of magnetization during a typical ${}^{75}$As NMR $T_{1}$ 
measurement at 100\,K.
(b) Temperature dependence of the ${}^{75}$As NMR relaxation rate 
$1/T_1$, measured at 7\,T. Below 100\,K, the relaxation (dominated by 
spin fluctuations) becomes increasingly faster and peaks \tcb{at 24\,K, 
i.e., well} above $T_{c}$ (see arrow). 
In the superconducting phase, the behavior is not exponential and $1/T_1$ 
saturates at a nonzero value, \tcb{here reflecting the presence of 
Pr$^{3+}$ magnetic ions}.}
\label{fig:invT1_T}
\end{figure}

The $^{75}$As spin-lattice relaxation times $T_1$ were evaluated from 
the magnetization recovery curves recorded 
\tcb{on the leftmost peak (at ca.\ 50.97 MHz)} at different temperatures. 
One of such curves, for $T=100$\,K, is shown in the inset of 
Fig.~\ref{fig:invT1_T}(a). The $T_1$ value for the central transition of 
a spin-3/2 $^{75}$As nucleus is obtained via \cite{McDowell1995}:  
\[ 
M_z(t) = M_z^0 \left[1-f\left(0.9e^{(-6t/T_1)^{\beta}} + 0.1e^{(-t/T_1)^{\beta}}\right) \right].\nonumber
\label{eq:T1_relax}
\] 
Here $M_z^0$ is the magnetization value at thermal equilibrium, $f$ 
reflects the efficiency of population inversion (ideally 2), and $\beta$ 
is a stretching exponent. Note that, while the reported data exhibit an 
almost ideal $\beta \sim 1$ value, at lower temperatures, due to intrinsic 
disorder induced by F-doping, $\beta$ decreases significantly [see 
Fig.~\ref{fig:invT1_T}(a)]. Under these circumstances, a low $\beta$ value 
indicates a wide distribution of relaxation rates.
Indeed, as the temperature is lowered, the inequivalence among the NMR 
sites increases. This implies a broader 
range of relaxation rates, in turn reflected in a decrease of $\beta$ from 1 (the ideal 
value in disorder-free metals) to ca.\ 0.65 at low temperature. The 
final upturn of $\beta$ close to 0\,K is not yet clear. 
Similar results have been found in the La-1111 family \cite{Shiroka2018} 
where, by systematically investigating samples across a large doping range, 
one can clearly correlate $\beta$ with the degree of disorder.

Our most important results are shown in Fig.~\ref{fig:invT1_T}(b), where 
we report the evolution of the spin-lattice relaxation rate with temperature. 
We recall that $1/T_1$ probes directly the fluctuations of the hyperfine 
fields at a nuclear site, since it is proportional to the $q$-summed value 
of $\chi''$, where $\chi''(q,\omega)$ is the imaginary part of the dynamical 
electronic susceptibility \cite{Abragam1978}.
The conspicuous 
increase of $1/T_1$ upon lowering temperature is clear evidence 
of the increasing importance of electronic spin fluctuations at low 
$T$. Such fluctuations are nothing but a residual of the original SDW 
magnetic order present in the undoped parent compound \cite{Ishida2011}. 
Indeed, it has been postulated that optimally doped samples strike a 
balance between the competing AF phase at lower dopings and the too weak 
fluctuations to sustain superconductivity at higher dopings. Our 
results confirm such a scenario also for the Pr-1111 case.

Let us now discuss the NMR relaxation data in more detail. As the 
temperature is lowered from room temperature, \tcr{first} we observe a 
gradual increase of fluctuations, peaking at a cusp-like maximum 
\tcb{above} 
$T_c$, followed by an abrupt decrease at low temperatures. 
If we compare our results with those obtained in similar \textit{Ln}1111 
compounds with different dopings \cite{Ishida2011,Shiroka2018}, it emerges 
that parent -- or very low doped -- compounds show a diverging behavior at 
$T_\mathrm{SDW}$, while overdoped samples show a very weak peak at $T_c$. 
Our case, closer to optimal doping (given the relatively high $T_c$), 
indicates that in PrFeAs(O,F) spin fluctuations are still dominant 
down to $T_c$. They would continue growing as the suppressed SDW is 
shifted towards 0\,K, but a change in the physics of the system close to 
$T_c$ clearly changes also their behavior. While in the undoped case, 
such event would be the opening of a SDW gap, in our case, the possibilities 
seem restricted to the intervening SC phase. However, since the $1/T_1$ peak 
occurs \tcb{at 25\,K, i.e., 10\,K} above $T_{c}$ \tcb{($\sim 15$\,K at 7\,T),} 
the most likely explanation for its occurrence might be given by the 
Bloembergen-Purcell-Pound 
\tcr{model \cite{Bloembergen1947,*Bloembergen1948},} as 
observed also in other under- or optimally doped La-1111 
compounds \cite{Hammerath2013}. Such model describes the behavior 
of the spin-lattice relaxation rate, $1/T_1$, under the influence of 
local fluctuating magnetic fields ~$h(t)$ and indeed predicts a peak 
in $1/T_1$ at the temperature where the effective correlation time of 
the spin fluctuations $\tau_c$ equals the inverse of the Larmor 
frequency $\omega_\mathrm{L}$. Considering the similarity of 
PrFeAs(O,F) with other \textit{Ln}1111 compounds, we expect the BPP 
model to apply also in our case. \tcb{Here, the particularly sharp 
cusp in $1/T_{1}$ might reflect the joint effect of the Fe$^{2+}$ 
and Pr$^{3+}$ spin fluctuations.}

Finally, we consider the $1/T_1$ behavior below $T_c$. \tcr{Given}  
the high quality of data, normally one could use them to study 
the superconducting gap and pairing. Unfortunately, close to 0\,K, 
the $1/T_1$ data converge at 0.3\,s$^{-1}$ and not at zero, as expected 
for a superconductor. In fact, deep in the SC phase, all electrons 
are bound into Cooper pairs, making the hyperfine interactions with 
the nuclei a very inefficient relaxation mechanism and driving 
the relaxation rate to zero. In our case, the finite value of $1/T_1(0)$ 
indicates that, at very low temperatures, other relaxation mechanisms 
are at play. \tcr{This excludes a possible use of the data collected 
in the SC phase and explains why here we limited our NMR study to the 
normal phase. Among the alternative relaxation mechanisms one could 
think of disorder-related relaxation channels (intrinsic to doping). 
However, a comparison with La-1111 results \cite{Hammerath2013,Shiroka2018}  
excludes it, since compounds with widely different dopings still exhibit a 
$1/T^{3}$ behavior at low $T$. On the other hand, the presence of a 
magnetic ion, such as Pr$^{3+}$, could well justify our results.
Indeed, an almost identical dependence of the ${}^{75}$As NMR relaxation 
rate vs.\ temperature is also found in the Ce-1111 case \cite{Rybicki2013}. 
We recall that Ce, Pr, and Nd have similar magnetic moments  
\tcb{(free-ion values of 2.54, 3.58, and 3.62\,$\mu_\mathrm{B}$, respectively)}, 
whose 
strong coupling with Fe spin fluctuations in the FeAs layer could explain 
our results, as well as the different low-$T$ behavior of relaxation 
compared with the nonmagnetic La-1111 case.}


\section{Conclusions}
By combining point-contact Andreev spectroscopy with nuclear magnetic 
resonance methods we investigated in detail the normal- and superconducting 
state properties of underdoped PrFeAs(O,F), a member of the 
\textit{Ln}1111 family.
Point-contact Andreev spectroscopy performed on single crystals 
provides evidence of the multiband/multigap nature of the 
PrFeAs(O,F) superconductivity. No indications of low-energy 
quasiparticles were found in the point-contact spectra, suggesting a 
fully gapped superconductor with no nodes. A small 
$\Delta_1 \simeq 3.5 \pm 0.5$\,meV gap was found not only to be very 
robust, but also to agree well with the results of microwave penetration 
depth and quasiparticle conductivity measurements \cite{hashimoto2009}. 
Additional structures, in the form of conductance maxima or shoulders, 
could be interpreted as being due to one \tcb{or possibly two larger 
gaps, whose amplitudes lie in the energy ranges $\simeq 6.0$--7.5\,meV 
(in agreement with optical conductivity measurements \cite{charnukha2018}) 
and $\simeq 8$--10\,meV}. The latter values are quite large and 
would correspond to gap ratios $2\Delta/k_\mathrm{B}T_c$ of the 
order of 9. \tcb{Moreover, we have shown that additional high-energy 
structures, ubiquitous in the conductance curves and not predicted by 
any BCS-based theory,} are the hallmark 
of \emph{strong coupling} between the electrons and spin fluctuations 
and can only be accounted for in the framework of a strong-coupling theory 
of superconductivity. Finally, magnetic resonance results in the normal 
phase provide clear evidence about the lack of any magnetic order in 
PrFeAs(O,F). Further, spin-lattice relaxation data suggest that this 
compound, similarly to other members of the 1111 family, hosts 
substantial electronic spin fluctuations \tcb{(here enhanced 
by the presence of Pr$^{3+}$ ions)}, which are expected to mediate the 
superconducting pairing.

\begin{acknowledgments} 
G.A.U.\ acknowledges support from the MEPhI Academic Excellence Project 
(Contract no.\ 02.a03.21.0005). This work was partly supported by the 
Swiss National Science Foundation (SNSF) through Grant no.\ 200021-169455.
\end{acknowledgments}

\appendix

\section{\label{app:PCAR_basics}Basics of the PCARS technique}
To ensure proper PCARS measurement conditions, the contact must be smaller than the electronic mean free path, so that the conduction is ballistic, no Joule dissipation occurs in the contact region, and the resistance of the contact 
largely exceeds the resistance of the normal bank \cite{daghero2010}. In these conditions, the voltage drop at the N/S interface practically coincides with the total potential difference between the electrodes, $V$, and the excess energy with which electrons are injected in the S side of the junction is just $eV$. Provided that there is no insulating layer at the sample surface, the conduction is dominated by the Andreev reflection \cite{Andreev,BTK}, even though the probability of quasiparticle tunnelling is not zero. The raw $dI/dV$-vs-$V$ spectrum 
already contains qualitative information on the number, amplitude, and symmetry of the gap(s). However, a more quantitative analysis can be made by fitting 
it with suitable models for the Andreev reflection at the N/S interface \cite{BTK,kashiwaya}.

To fabricate the contacts, we used the so-called ``soft'' technique, 
widely described elsewhere \cite{daghero2010,zhigadlo2018}. In a few 
words, we stretch a thin Au wire over the crystal, until it touches the 
surface in a single point. Typically, ballistic contacts have resistances 
of a few tens of Ohms (even though the actual value depends on the 
properties of the sample, namely on its normal-state resistivity). 
This type of contacts can be mechanically unstable, especially during cooling/heating, because of the different thermal coefficients of the 
materials. Thus, in some cases, we used a drop of conducting Ag glue 
to improve the stability. Independent of the presence of Ag glue, the 
actual contact must be thought of as a parallel of nanoscopic contacts 
between a normal metal and a superconductor.

\section{\label{app:PCAR_theory}A strong-coupling model for the superconductivity in Pr-1111}
As already mentioned, one can account for the presence of electron-boson 
structures only by using a strong-coupling extension of the BCS theory, 
i.e., the Eliashberg theory. Based on the similarity with other 
electron-doped Fe-based superconductors, we assume Pr-1111 to be 
described by an effective $s\pm$-wave three-band model \cite{mazin_schmalian,Eliashberg,Chubukov,Manske}, with one hole-like band 
centered at $\Gamma$ and two electron-like Fermi surface sheets at 
the corners of the Brillouin zone. This model is described in detail 
elsewhere \cite{Umma1,dagheroRopp2011}.

To calculate the SC gaps and the critical temperature in this model, 
one has to solve six coupled equations for the complex order parameters $\Delta_{i}(i\omega_{n})$ and the renormalization functions $Z_{i}(i\omega_{n})$, where $i=1,2,3$ is the band index and $\omega_{n}$ are the Matsubara frequencies. The frequency (energy) dependence of the order parameters, normally ignored in the BCS theory, is 
here the key factor that accounts for the presence of the electron-boson coupling features. There are many input parameters, including: i) nine electron-phonon spectral functions, $\alpha^{2}_{ij}F^{ph}(\Omega)$; ii) nine electron-boson (spin fluctuactions) spectral functions, $\alpha^{2}_{ij}F^{sf}(\Omega)$; iii) nine elements of the Coulomb pseudopotential matrix, $\mu_{ij}^{*}(\omega_{c})$. 
To a first approximation, we neglect the disorder, thus assuming that 
all the scattering rates (from either magnetic or nonmagnetic impurities) 
are zero. To further simplify the problem, some additional assumptions 
can be made, shown to be valid in the case of iron pnictides \cite{Umma1,Umma2,Umma3}. In particular, we know that phonons do not contribute significantly to the (dominant) interband coupling \cite{mazin_schmalian}  
\begin{figure}[th]
	\includegraphics[keepaspectratio, width=\columnwidth]{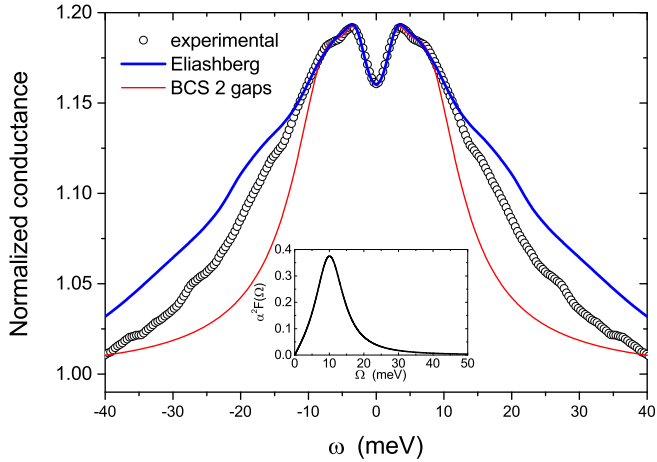}
	\caption{Open circles: experimental data reported in Fig.~\ref{fig:lowTfit}(b). Blue line: calculated curve after inserting into the three-band BTK-like model the energy-dependent order parameters, calculated by solving the Eliashberg equations. The red line is a fit of the theoretical curve with a two-band BCS-based BTK model (with constant gaps), which reproduces the same gap values as in Fig.~\ref{fig:lowTfit}(b) (red curve). The inset shows the electron-boson spectral function for the antiferromagnetic spin fluctuactions, here normalized to have $\lambda=1$.}
	\label{fig:Eliashberg}
\end{figure}
(i.e., $\lambda^{ph}_{ij}\approx 0$) and that the total electron-phonon coupling constant is small \cite{Boeri2}. Hence, we can simply neglect the phonon contribution and assume $\lambda^{ph}_{ii}=0$. Spin fluctuations, instead, 
are known to provide mostly the \emph{inter}band coupling between the 
hole- and electron bands, so we can assume $\lambda^{sf}_{ii}=0$. Finally, following Ref.~\cite{Mazincoulomb}, we will assume that the Coulomb pseudopotential matrix is identically zero, i.e., $\mu^{*}_{ii}(\omega\ped{c})=\mu^{*}_{ij}(\omega\ped{c})=0$.
With these approximations, the electron-boson coupling-constant 
matrix $\lambda_{ij}$ becomes \cite{Umma1,Umma2,Umma3,mazin_schmalian}:
\begin{equation}
\vspace{2mm} %
\lambda_{ij}= \left (
\begin{array}{ccc}
0                 &         \lambda^{sf}_{12}                  &               \lambda^{sf}_{13}            \\
0                &               \lambda^{sf}_{21}=\lambda^{sf}_{12}\nu_{12}               &                0            \\
\lambda^{sf}_{31}=\lambda^{sf}_{13}\nu_{13} &  0  & 0 \\
\end{array}
\right ), \label{eq:matrix}
\end{equation}
where $\nu_{ij}=N_{i}(0)/N_{j}(0)$, with $N_{i}(0)$ the normal
density of states at the Fermi level for the $i$-th band. The electron-boson coupling constants are defined through the Eliashberg functions 
$\alpha^2_{ij}F_{ij}^{sf}(\Omega)$. Following Refs.~\cite{Umma1,Umma2,Umma3} we choose these functions to have a Lorentzian shape, with a maximum at the energy $\Omega_0$ and half-width $\Omega_{0}/2$. $\Omega_0$ is the characteristic energy of the mediating boson, which corresponds to the characteristic energy of the spin resonance \cite{Inosov} and is related to the critical temperature by the empirical law $\Omega_0=2 T_c /5$ that has been demonstrated to hold, at least approximately, for many iron pnictides \cite{Paglione}.

The factors $\nu_{ij}$ that enter the definition of $\lambda_{ij}$ can 
normally be determined from the band-structure calculations, unfortunately not 
available for the Pr-1111 case. However, since our aim is just to show 
that the wide shoulders observed in our datasets (we will refer in particular to the data in Fig.~\ref{fig:lowTfit}b) are due to electron-boson coupling, 
we may use the values employed in Co-doped Ba-122 \cite{Umma3} (because 
this also is an electron-doped Fe-based superconductor with a similar 
$T_c$), i.e., $\nu_{12}=1.12$ and $\nu_{13}=4.50$. Hence, only two free parameters remain, $\lambda_{12}$ and $\lambda_{13}$, that need to be fixed in order to reproduce the experimental $T_c$ and the experimental gaps. 
We find $\lambda_{12}=0.7$ and $\lambda_{13}=1.8$, giving a total 
coupling constant $\lambda_{t}=2.374$.

Owing to the $s\pm$ symmetry, the order parameter has opposite signs on the hole-like and the electron-like Fermi surface sheets. The low-temperature values of the gaps turn out to be $\Delta_{h}=6.33$\,meV, $\Delta_{e1}=3.35$\,meV and $\Delta_{e2}=8.56$\,meV, where the subscripts $h$ and $e$ refer to the hole-like 
and electron-like Fermi surfaces, respectively. The gap values agree 
very well with the gap distribution shown in Fig.~\ref{fig:gaps}. The calculated critical temperature is $T_{c}=28.6 $\,K. This value is  slighty larger than the onset of the superconducting transition (see Fig.~\ref{fig:magn_res}). However, since the coupling is of electronic origin, there is a feedback effect of the SC condensate on the spin-fluctuation spectrum \cite{Chubukov,Manske}. 
Taking this effect into account, as we already did in Ref.~\cite{Umma3}, 
the critical temperature turns out to be $T_{c}=22.86$\,K.
Once the order parameters as a function of energy are known, they can 
be inserted into the equations for the Andreev reflection (i.e., in the three-band version of the BTK model) and the conductance curve can be calculated. The BTK model contains, in addition to the gap amplitudes, the relative weights of the bands, the barrier parameters and the broadening parameters. We chose the values of these parameters in order to obtain a curve similar to that 
in Fig.~\ref{fig:lowTfit}(b). In particular, we took $Z_{h}=Z_{e1}=Z_{e2}=0.33$, $\Gamma_{h}=2.95$\,meV, $\Gamma_{e1}=1.45$\,meV, $\Gamma_{e2}=4.05$\,meV, 
$w_{h}=0.20$, $w_{e1}=0.25$, $w_{e2}=0.55$. The resulting curve 
is shown in Fig.~\ref{fig:Eliashberg} with a blue line. Clearly, 
accounting for the energy dependence of the order parameters gives rise to very wide (in energy) and very high (in amplitude) shoulders that resemble very closely those of the experimental data (actually, a proper normalization could easily reproduce an experimental spectrum with the same shape), \tcb{but does not affect the low-energy part of the spectrum, where the gap-related features show up. Indeed,} the fit of the theoretical curve with the same two-band BTK model we 
used to fit the experimental data would have given again $\Delta_{1}=3.3$\,meV and $\Delta_{2}=8.5$\,meV.

%

\end{document}